\documentclass[prl,twocolumn,superscriptaddress,amsmath,amssymb,nofootinbib]{revtex4-1}

\usepackage{braket,mathtools}
\usepackage{amsmath,amssymb,amsthm}
\usepackage{upgreek}
\usepackage[unicode=true]{hyperref}
\usepackage{subfigure,bbm,times}
\usepackage{xcolor}
\usepackage[T1]{fontenc}
\usepackage[normalem]{ulem}

\renewcommand{\tilde}{~}
\newcommand{\id}{1\!\!1}

\newcommand{\trace}[1]{\text{Tr}\left[#1\right]}
\newcommand{\sign}[1]{\text{sgn}\left(#1\right)}
\newcommand{\psin}{\ket{\psi_{\text{in}}}}

\newcommand{\cypermut}{\pi_{(1,2,\dots,N)}}
\newcommand{\projen}{P_{E_{(-1)^{N-1}}(\cypermut)}}

\newcommand{\rin}{\rho_\text{in}}
\newcommand{\rdep}{\rho_\text{dep}}
\newcommand{\supp}[1]{\text{supp}(#1)}

\begin{document}

    \title{Robust generation of $N$-partite $N$-level singlet states by identical particle interferometry}

    \author{Matteo Piccolini}
        \affiliation{Dipartimento di Ingegneria, Universit\`{a} di Palermo, Viale delle Scienze, 90128 Palermo, Italy}

    \author{Marcin Karczewski}
        \email{marcin.karczewski@ug.edu.pl}
        \affiliation{International Centre for Theory of Quantum Technologies, University of Gda\'{n}sk, 80-308 Gda\'{n}sk, Poland}

    \author{Andreas Winter}
        \email{andreas.winter@uab.cat}
        \affiliation{ICREA {\&} Grup d'Informaci\'o Qu\`antica, Departament de F\'{\i}sica, Universitat Aut\`{o}noma de Barcelona, 08193 Bellaterra (Barcelona), Spain}
        \affiliation{Institute for Advanced Study, Technische Universit\"at M\"unchen, Lichtenbergstra{\ss}e 2a, 85748 Garching, Germany}
        \affiliation{QUIRCK --- Quantum Information Independent Research Centre Kessenich, Gerhard-Samuel-Stra{\ss}e 14, 53129 Bonn, Germany}

    \author{Rosario Lo Franco}
	\email{rosario.lofranco@unipa.it}
	\affiliation{Dipartimento di Ingegneria, Universit\`{a} di Palermo, Viale delle Scienze, 90128 Palermo, Italy}

    \begin{abstract}
        We propose an interferometric scheme for generating the totally antisymmetric state of $N$ identical bosons with $N$ internal levels (generalized singlet). This state is a resource for various problems with dramatic quantum advantage. The procedure uses a sequence of Fourier multi-ports, combined with coincidence measurements filtering the results. Successful preparation of the generalized singlet is confirmed when the $N$ particles of the input state stay separate (anti-bunch) on each multiport. The scheme is robust to local lossless noise and works even with a totally mixed input state.
    \end{abstract}

    \maketitle

    \textit{Introduction.---}
        Manipulation of entangled states is necessary to fully access the advantages of quantum technologies. For this reason, great attention has been dedicated to classes of entangled states proven to be useful for quantum information tasks, ranging from the simplest 2-qubit Bell states\tilde\cite{nielsen2010quantum}, to more complex classes of many-body systems such as W states\tilde\cite{dur2000three}, GHZ states\tilde\cite{greenberger1989going,mermin1990quantum}, NOON states\tilde\cite{lee2002quantum}, Dicke states\tilde\cite{dicke1954coherence}, and many more\tilde\cite{polino2020photonic}.
        
        Crucially, quantum correlations characterizing such states must be protected from the detrimental action of external noise to allow for their real-world exploitation. A plethora of techniques has been suggested to achieve this goal, including decoherence-free subspaces\tilde\cite{zanardi1997noiseless, lidar1998decoherence}, structured environments with memory effects\tilde\cite{mazzola_2009,bellomo_2008,lo_franco_2013,Xu_2010,bylicka_2014,man_2015,tan_2010,tong_2010,breuer_2016_colloquium,man_2015_pra}, quantum error correction codes\tilde\cite{preskill_1998,knill2005quantum, shor_1995,steane_1996}, dynamical decoupling and control techniques\tilde\cite{Viola1998,viola2005random,darrigo_2014_aop,franco2014preserving,orieux_2015,facchi_2004,lo_franco_2012_pra, xu_2013,damodarakurup_2009,cuevas_2017}, quantum repeaters\tilde\cite{briegel1998quantum,duan2001long,munro2015inside}, distillation protocols\tilde\cite{bennett1996purification,horodecki1997inseparable,horodecki1998mixed,horodecki2001distillation,kwiat_2001,dong_2008}, and interferometric effects in systems of identical particles\tilde\cite{indistentanglprotection,nosrati2023indistinguishabilityassisted,Piccolini_2021_entropy,piccolini2021opensys,piccolini2023asymptotically,piccolini2023robust}.
        These techniques can be applied to physical systems featuring a wide range of inherent fragility to environmental noise. In particular, photons have a long coherence time, making them suitable for long-distance communications between remote parties\tilde\cite{gisin2007quantum,xu2020secure}. However, there is a trade-off: as photons interact little with each other directly, entangling them requires alternative methods such as nonlinear multiphoton generation techniques (such as SPDC\tilde\cite{kwiat1995new} and four-wave mixing\tilde\cite{li2005optical,takesue2004generation}), heralding processes\tilde\cite{knill2001scheme,zhang2008demonstration,varnava2008good,gubarev2020improved,chin2022shortcut,chin2023graphs}, or postselected measurements of identical photons spatially overlapping over detection regions\tilde\cite{slocc,experimentalslocc,sun2022activation}.
        
        These techniques are typically employed in long-range communication scenarios, in which the resource states prepared by the sender require protection during their propagation through noisy environments.
        In Refs.\tilde\cite{piccolini2023asymptotically,piccolini2023robust}, the authors shift this viewpoint by proposing a protocol where the entangled resource is prepared by the receiver after the environmental noise has affected the system. To this end, they devise a scheme to prepare maximally entangled states of two identical qubits which probabilistically succeeds regardless of the initial state, that is, regardless of local particle-preserving noise previously acting on the system. This goal is achieved by locally injecting white noise on the two qubits, which resets the system to the maximally mixed state. The various components of the mixture interfere differently under the action of a beam splitter, with bosonic particles in the singlet state staying separate (anti-bunching) and the ones in symmetric states grouping together (bunching). This effect can then be exploited to postselect a pure Bell singlet state of two identical bosons with coincidence measurements.

        In the present work, we extend this protocol to $N$ identical bosons with $N$ internal levels and devise a scheme to generate an $N$-partite singlet state from a maximally mixed state having one particle per mode. The multipartite singlet state, which is antisymmetric under the exchange of any pair of particles, can be exploited to solve 
        communication tasks which have no known classical solution\tilde\cite{cabello2002n,cabello2003solving,cabello2003supersinglets} and certify the non-projective character of measurements\tilde\cite{ma2023randomness}. Furthermore, entangled states of spatially distinguishable particles can be used to simulate particle statistics of different types, a phenomenon which has been shown to obey monogamous relations: a totally antisymmetric state of $N$ distinct bosons can thus provide a useful testing ground to study the properties of $N$ spatially indistinguishable fermions\tilde\cite{karczewski2018monogamy}. Finally, generalized singlet states are invariant under global rotations of the internal levels and are characterized by a zero variance of the related pseudospin operator $J^2$. This property makes them potentially useful in quantum metrology, where they can be used to probe local fields with enhanced performances\tilde\cite{urizar2013macroscopic}.

        The usefulness of generalized singlet states is hindered by the difficulty of obtaining them. To achieve this, a method based on a sequence of quantum nondemolition (QND) measurements has been proposed in\tilde\cite{toth2010generation}. This technique, already implemented with both cold\tilde\cite{behbood2014generation} and hot\tilde\cite{kong2020measurement} atomic ensembles, involves postselection and allows for the preparation of a state approximating the generalized singlet. However, it does not lead to an exact, pure generalized singlet state, not even when endowed with a feedback mechanism implementing corrections between the measurements\tilde\cite{behbood2013feedback}.
        
        In contrast to this, the technique we propose here allows for the probabilistic preparation of exact multipartite singlet states. To do so, we relate the behavior of $N$ identical particles injected in an $N$-port interferometer to their symmetries, as dictated by the general suppression law reported in Ref.\tilde\cite{dittel2018totally}. Subsequently, we use the obtained insights to devise a scheme composed of a sequence of Fourier $2,\ldots,N-1,N$-port interferometers interlaced with QND coincidence measurements performed on the related output modes.
        As in Refs.\tilde\cite{piccolini2023robust,piccolini2023asymptotically}, the maximally mixed initial state guarantees that our procedure is robust under the action of local noise acting on the $N$ particles. Finally, we propose an implementation that employs postselection and a specific initial state to prepare an $N=3$ generalized singlet state without QND measurements.

    \textit{Generalized singlet state.---}
        The goal of this work is to design an interferometric procedure to prepare the generalized singlet state of $N$ spatially separated bosons with $N$ internal levels
        \begin{equation}
        \label{singlet}
            \ket{A_N}:=\frac{1}{\sqrt{N!}}\sum_{\pi\in S_N} \prod_{i=1}^N\sign{\pi}a^\dagger_{i,\pi(i)}\ket{0},
        \end{equation}
        where $\sign{\pi}$ is the sign of the permutation $\pi$ from symmetric group $S_N$, and $a^\dagger_{\ell,m}$ denotes the operator creating a particle with internal state $m$ in spatial mode $\ell$. 
        For example, $\ket{A_2}=(\ket{0,1}-\ket{1,0})/\sqrt{2}$ is the ordinary Bell singlet state of two qubits, whereas for three qutrits we have $\ket{A_3}=(\ket{0,1,2}-\ket{0,2,1}-\ket{1,0,2}+\ket{1,2,0}+\ket{2,0,1}-\ket{2,1,0})/\sqrt{6}$. The interest in this class of states stems from their rotational invariance, leading to applications in quantum protocols \tilde\cite{cabello2002n}, and total antisymmetry, which brings these bosonic states as close to fermionic properties as possible \tilde\cite{karczewski2018monogamy}. 
        
        The systematic classification of the types of symmetries of $N$ particles with $d$ internal levels can be achieved with representation theory\tilde\cite{fulton1997young}. One of its basic results states that the space of totally antisymmetric states of $N$ constituents with $N$ internal levels is one-dimensional, that is, $\ket{A_N}$ is the unique totally antisymmetric state of the considered system.

    \textit{Suppression law for anti-bunching.---}
     In order to prepare the generalized singlet state $\ket{A_N}$ given in Eq.\tilde\eqref{singlet}, we consider the transformation of the input state under a Fourier $N$-port given by 
             \begin{equation}
             \label{creation}
        b^\dagger_{k,m}=\sum_{\ell=1}^N (U_N)_{k,\ell}\,a^\dagger_{\ell,m},    
        \end{equation}
        where $b^\dagger$ denotes the creation operator for the output mode and the matrix $U_N$ is given by
           \begin{equation}
        \label{fourier}
            U_N=\frac{1}{\sqrt N}
            \begin{pmatrix}
                1 & 1 & 1 & \dots & 1\\
                1 & \omega & \omega^2 & \dots & \omega^{N-1}\\
                1 & \omega^2 & \omega^4 & \dots & \omega^{2(N-1)}\\
                \vdots & \vdots & \vdots & \vdots & \vdots \\
                1 & \omega^{N-1} & \omega^{2(N-1)} & \dots & \omega^{(N-1)(N-1)}
            \end{pmatrix}
        \end{equation}
        for $\omega=e^{2\pi i/N}$. Since the state $\ket{A_N}$ has a single particle in each spatial mode, we would like to pin down the conditions that an $N$-particle input state must satisfy in order to anti-bunch on the Fourier $N$-port.
        
        Conveniently, in Ref.\tilde\cite{dittel2018totally} Dittel \textit{et al}. provide a suppression law characterizing prohibited outcomes in interferometric experiments for a class of multiports including Fourier $N$-ports. Their results imply that the eigenstates $\ket{\varphi}$ of the cyclic permutation $\cypermut$ that can anti-bunch when transformed by $U_N$ need to obey (see Supplemental Note I \cite{[{See Supplemental Material at }]supp})
        \begin{equation}
        \label{symmetrycond}
        \cypermut \ket{\varphi}=(-1)^{N-1}\ket{\varphi}.
        \end{equation}

        To use this insight, let us note that a generic input state $\psin$ of $N$ particles with $N$ internal levels can always be decomposed in the $N$-particle eigenbasis of $\cypermut$. Then, the condition Eq.\tilde\eqref{symmetrycond} rules that $\psin$ can anti-bunch when transformed by $U_N$ only if its projection onto the $(-1)^{N-1}$-eigenspace of $\cypermut$ is nonzero.
        We denote this eigenspace $E_{(-1)^{N-1}}(\cypermut)$ and define the related projection operator
        \begin{equation}
        \label{enprojector}
            \projen:=\frac{1}{N}\sum_{k=1}^{N} \left[(-1)^{N-1}\,\cypermut\right]^k.
        \end{equation}
        
        It is clear now that \textit{a necessary condition for a generic $N$-boson input state $\rho_\textup{in}$ to anti-bunch on a Fourier $N$-port reads}
        \begin{equation}
        \label{anti-bunchcondition}
            \trace{\rho_\textup{in}\projen}\neq 0.
        \end{equation}

        \textit{Implementing the eigenspace projector.---}
        Condition given by Eq.\tilde\eqref{anti-bunchcondition} can also be interpreted as an operational recipe for implementing a projection into the eigenspace $E_{(-1)^{N-1}}(\cypermut)$. It consists in casting an input state composed of $N$ particles, one in each spatial mode, on a Fourier $N$-port followed by performing a coincidence measurement on the output modes. In particular, this measurement can be realized by means of $N$ quantum non-demolition single particle detectors filtering out non-coincident detections, effectively implementing the operator $C_N=\sum_{\sigma_1,\dots,\sigma_N=0}^{N-1}\ket{\sigma_1,\dots,\sigma_N}\!\bra{\sigma_1,\dots,\sigma_N}$. This constitutes the basic step of our protocol.

 \textit{Extracting the singlet.---} Let us now consider a sequence of the above steps with the size of the Fourier multiport increasing from 2 to $N$, defining $M_k:=\prod_{j=k}^2 C_jU_j$ (notice that the index in the product decreases to reflect the order of the operations). We are going to show that for an input state with a single $d$-level particle in each of $N$ modes we have
 \begin{align}
 \label{induction}
 M_N=e^{i \phi_N} \prod_{j=N}^2 P_{E_{\left((-1)^{j-1}\right)}(\pi_{(1,\ldots,j)})},
 \end{align}
 where $\phi_N$ is an irrelevant global phase.
 By direct calculation one can verify that Eq.~(\ref{induction}) is satisfied for $N=2$. Suppose now that it holds for any $k<N$, with $N>2$. Since Eq.\tilde\eqref{anti-bunchcondition} provides a necessary condition for anti-bunching, we have $\supp{C_N U_N}\subseteq\text{Im}(\projen)$.  Thus we have 
 \begin{align}
     M_N&=  C_N U_N P_{E_{\left((-1)^{N-1}\right)}(\pi_{(1,\ldots,N)})} \prod_{j=N-1}^{2} C_jU_j\nonumber\\&=e^{i \phi_{N-1}} C_N U_N  \prod_{j=N}^2  P_{E_{\left((-1)^{j-1}\right)}(\pi_{(1,\ldots,j)})}.
 \end{align}
  But it can be shown (see Supplemental Note II\tilde\cite{supp}) that
               \begin{equation}
           \prod_{j=N}^2 P_{E_{\left((-1)^{j-1}\right)}(\pi_{(1,\ldots,j)})}=\frac{1}{N!}\sum_{\pi \in S_N} \text{sgn}(\pi)\,\pi=P_{A_N^d},
            \end{equation}
        where $P_{A_N^d}$ is the projector onto the totally antisymmetric subspace of $N$ particles with $d$ internal levels.
        Therefore $M_N = e^{i \phi_{N-1}} C_N U_N P_{A_N^d}$. It remains to be shown that any state from the totally antisymmetric subspace is invariant under Fourier multiport $U_N$ and coincident detection $C_N$.
 For the sake of simplicity we restrict our attention to $d=N$ (see Supplemental Note III\tilde\cite{supp} for the general case). The totally antisymmetric space is then spanned by $ |A_N\rangle$. From Eq.~(\ref{creation}) and Eq.~(\ref{singlet}) it follows that
\begin{equation}
    U_N|A_N\rangle = \frac{1}{\sqrt{N!}}\det (U_N \mathsf{A})\, |0\rangle,
\end{equation}
where 
\begin{equation}
\mathsf{A}=\begin{pmatrix}
     a^\dagger_{1,1} & a^\dagger_{1,2} & \ldots & a^\dagger_{1,N}\\ 
     a^\dagger_{2,1} & a^\dagger_{2,2} & \ldots & a^\dagger_{2,N}\\
     \vdots&\vdots &\vdots &\vdots\\
     a^\dagger_{N,1} & a^\dagger_{N,2} &\ldots  & a^\dagger_{N,N}.
\end{pmatrix}
\end{equation}
But as $\det (U_N \mathsf{A})=\det U_N \det \mathsf{A} = (-1)^{N+1} \det \mathsf{A}$ we get that $ U_N|A_N\rangle=(-1)^{N+1}|A_N\rangle$. Clearly the global phase shift is irrelevant, and the fact that $|A_N\rangle$ has a single particle in each mode ensures that it is not affected by the coincidence measurement $C_N$.

\begin{figure}
            \centering
            \includegraphics{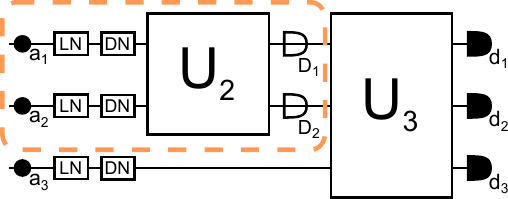}
            \caption{Procedure for the preparation of the $N=3$ bosonic generalized singlet state $\ket{A_3}$. Three identical bosons localized on distinct modes are subjected to arbitrary local noise (LN) and subsequently depolarized (DN). Later on, two of them are cast onto a beam splitter $U_2$. Two single-particle non-absorbing detectors perform a coincidence measurement on the output modes, selecting only the anti-bunched results. The three particles are then injected into a tritter $U_3$. Three single-particle detectors perform a final coincidence measurement on the output, collecting only the anti-bunched states. The last step can be done either with QND detectors or with absorbing detectors via postselection. Part of the scheme enclosed in a dashed box can be replaced with a heralded generation of the singlet state $\ket{A_2}$.}
            \label{fig:sketch}
        \end{figure}

        This shows that \textit{the generalized singlet state of $N$ identical bosons with $N$ levels can be probabilistically distilled from an arbitrary initial state $\rin$ with a single particle per mode by acting on it with a sequence of Fourier $2-,\dots,N$-ports and selecting only the results which anti-bunch at every step} (see Fig.\tilde\ref{fig:sketch} for a pictorial representation of the setup for $N=3$).
        The procedure we just described can be seen as a filtering scheme where the generalized singlet component of the input state $\rin$ is probabilistically distilled. Its success probability $p_s=\trace{\ket{A_N}\!\bra{A_N}\rin}$ depends on the overlap of the initial state with the generalized singlet, in particular being null when there is none.
        
        It should be stressed that the coincidence measurements $C_j$ must be nondemolitive, as the particles emerging from a Fourier multiport are later cast onto the next one. Such measurements can be implemented with nonabsorbing detectors\tilde\cite{braginsky1996quantum,roch1992quantum,unnikrishnan2015quantum}. This requirement does not hold for the last measurement (following the Fourier N-port), which can be realized using standard single particle detectors in a postselected implementation\tilde\cite{experimentalslocc,sun2022activation,wang2022proof,piccolini2023asymptotically,piccolini2023robust}.
      
        \textit{Robust generalized singlet preparation.---}
        As previously stated, we want to distill the generalized singlet state in a way which is robust to the action of lossless local noise acting on the initial state. To do so, we start with an arbitrary state $\rho_N$ of $N$ identical bosons with $N$ internal levels occupying one spatial mode each. Following the idea introduced in Ref.\tilde\cite{piccolini2023asymptotically,piccolini2023robust}, we act on each particle with local externally-activated depolarizing noise, obtaining the $N$-body maximally mixed state $\rdep=\bigotimes_{j=1}^{N}\rho_j$, where $\rho_j$ is the Werner state $\rho_j=\frac{1}{N}\sum_{k=0}^{N-1}\ket{k}_j\!\bra{k}_j$ of the particle in the $j^\text{th}$ mode with internal level $k$. This operation has a double role: first of all, it resets the system to the known state $\rin=\rdep$, thus making the obtained result independent of the original state $\rho_N$, of the characteristics of the noisy environments acting on the constituents prior to the depolarization, and on the interaction time between them. Secondly, it ensures that the state $\rin$ injected in the setup has a nonzero overlap with the generalized singlet, guaranteeing that the probability of extracting $\ket{A_N}$ is non-zero. Indeed, $\rdep$ can always be expressed in a diagonal form on a $N^N$-dimensional orthonormal basis which includes $\ket{A_N}$. Therefore, the proposed preparation technique is deemed to be robust, although it succeeds with probability
        \begin{equation}
        \label{wernerprob}
            p_s=\trace{\ket{A_N}\!\bra{A_N}\rdep}=1/N^N.
        \end{equation}

        \textit{Alternative realizations with specific initial states.---}
        In the suggested implementation, robustness to noise is obtained in exchange for a low success rate and the requirement of QND measurements. Nonetheless, these two drawbacks can be mitigated when a free choice of the initial state is allowed. This is the case, for example, of a preparation occurring immediately after the state initialization, or when the noise affecting the system between the source and the implementation of our scheme is negligible.
        In these scenarios, for instance, the preparation of $N$ spatially separated bosons in the product state $\ket{0,1,\dots,N-1}$ would guarantee an enhanced success probability of $p_s=1/N!$.
        
        The possibility to choose the initial state also allows to avoid relying on nonabsorbing detectors in specific scenarios, opening the path for realistic experimental implementations. Consider, for example, $N=3$ qutrits in the initial state $\psin=\ket{A_2}\otimes\ket{2}$. The implementation of the Fourier 2-port (beam splitter) can now be avoided, reducing our setup to a single tritter: indeed, it can be easily checked that $C_3U_3\psin=\ket{A_3}$, so that the generalized singlet state is distilled with probability $p_s=\lvert\braket{A_3|\psi_\text{in}}\rvert^2=1/3$.
        Similarly, we can allow for the third qutrit to be depolarized as in the robust approach (see Figure\tilde\ref{fig:sketch}), obtaining $\rin=\ket{A_2}\!\bra{A_2}\otimes\frac{1}{3}\big(\sum_{k=0}^{2}\ket{k}\!\bra{k}\big)$ and preparing $\ket{A_3}$ with probability $p_s=\trace{\ket{A_3}\!\bra{A_3}\rin}=1/9$.
        Crucially, the QND measurement $C_3$ can be substituted in both cases by a postselection carried out with standard single-particle detectors. The preparation of such initial states only requires the ability to entangle $2$ qutrits in a Bell singlet-like state and to eventually depolarize a third one, a challenge which could be tackled, for example, with frequency-bin manipulation techniques\tilde\cite{clementi2023programmable}, thus making the preparation of the bosonic generalized singlet of $3$ qutrits an experimentally feasible task.

        \textit{Conclusions.---}
        We have introduced a theoretical protocol to probabilistically prepare the totally antisymmetric state $\ket{A_N}$ of $N$ distinct bosons with $N$ internal levels. This state finds potential applications in quantum information protocols \tilde\cite{cabello2002n,cabello2003solving,cabello2003supersinglets}, simulating systems of fermionic indistinguishable particles\tilde\cite{karczewski2018monogamy}, certifying the non-projective character of measurements\tilde\cite{ma2023randomness}, and in quantum metrology\tilde\cite{urizar2013macroscopic}.
        
        The scheme, which generalizes the one devised in Ref.\tilde\cite{piccolini2023asymptotically,piccolini2023robust} to many-body systems, employs a sequence of Fourier multiports with the number of ports ranging from $2$ to $N$, interlaced with coincidence measurements distilling the results where one constituent per mode is found. The measurements, which have to be insensitive to the internal degree of freedom, must preserve the detected particles and are thus required to be nondemolitive. This does not hold for the last coincidence count, which can be deferred and realized with standard absorbing detectors via postselection\tilde\cite{experimentalslocc,sun2022activation,wang2022proof}. We stress that the emergence of the generalized singlet from the proposed setup is merely due to the interference effects between the identical constituents generated by the Fourier multiports, as discussed in Ref.\tilde\cite{piccolini2023asymptotically,piccolini2023robust}.
        Therefore, our work supports the perspective of identicality as a potential quantum resource. 
        
        The success probability depends on the overlap between the initial state and the generalized singlet, as the latter has been shown to be the only state to satisfy the necessary condition to anti-bunch under the proposed setup. This property has been used to propose a feasible scheme where the $N$ particles are initially externally depolarized, leading to a maximally mixed state which always has nonzero overlap with the generalized singlet. This strategy also  allows one to ignore the previous history of the system, including the initially prepared state and the eventual local interaction of the $N$ particles with lossless noisy environments. This feature enables our scheme to successfully prepare the generalized singlet state even when the setup is implemented far from the particles source, assuming no particle losses. In this sense, the proposed protocol is robust against local noise acting prior to the externally-induced depolarization.
        
        With the suggested realization, the success probability is found to scale as $1/N^N$. Nonetheless, alternative initial states can be employed to provide higher success rates when the presence of noise is low enough to avoid resorting to the external depolarization. This occurs, for example, when our scheme can be applied immediately after the preparation of the initial state.
        Although our findings further point out the relevance QND detectors might have for quantum information protocols, we have shown that specific experimentally-achievable initial states can be used to obtain generalized singlet states without relying on nonabsorbing detectors. 
        Looking for initial states exploitable to generate high-dimensional generalized singlet states with current technology is surely a direction which is worth of further investigation. 
        Moreover, in Ref.\tilde\cite{piccolini2023robust} the proposed scheme was shown to distill pure $N=2$ NOON states when applied to two identical fermions. It would thus be interesting to work out its generalization to multipartite fermionic systems.

    \begin{acknowledgments}
    R.L.F. acknowledges support from European Union -- NextGenerationEU -- grant MUR D.M. 737/2021 -- research project ``IRISQ''. M. K. acknowlegdes support from Foundation for Polish Science (IRAP project, ICTQT, contract no.2018/MAB/5, co-financed by EU within Smart Growth Operational Programme).
    AW is supported by the European Commission QuantERA grant ExTRaQT (Spanish MICIN project PCI2022-132965), by the Spanish MCIN (project PID2022-141283NB-I00) with the support of FEDER funds, by the Spanish MCIN with funding from European Union NextGenerationEU (PRTR-C17.I1) and the Generalitat de Catalunya, and by the Ministry  of Economic Affairs and Digital Transformation of the Spanish  Government through the QUANTUM ENIA project call - Quantum Spain project, with funding from European Union NextGenerationEU within the framework of the "Digital Spain 2026 Agenda". Furthermore by the Alexander von Humboldt Foundation and the Institute for Advanced Study of the Technical University Munich.
    M.P. and M.K. equally contributed to this work.
    \end{acknowledgments}


%

\renewcommand{\theequation}{S\arabic{equation}}
\setcounter{equation}{0}

    \section{Supplemental Note I}
    In this section we briefly review the suppression law which led to the equation
    \begin{equation}
    \label{symmetrycond_supp}
        \cypermut \ket{\varphi}=(-1)^{N-1}\ket{\varphi}.
    \end{equation}
    reported in the main text.
    
    In [Phys. Rev. A 97, 062116 (2018)] the authors derive a general suppression law for any pure initial state $\psin$ of identical bosons distributed over $n$ spatial modes subjected to a unitary mode-mixing evolution. The particles are further characterized by an internal degree of freedom $\ket{I}$, with $I\in\{1,2,\dots,d\}$. Such a suppression law is ultimately found to be strictly related to the permutation symmetries of the initial state $\psin$.
    In particular, $\psin$ is characterized by the \textit{mode occupation list} $\Vec{r}=(r_1,\dots,r_n)$ describing the number $r_j$ of particles occupying the $j^\text{th}$ mode. To such an input configuration is associated the \textit{mode assignment list} $\Vec{d}(\Vec{r})=(d_1(\Vec{r}),\dots,d_N(\Vec{r}))$, where $N$ is the total number of particles and $d_\alpha(\Vec{r})\in\{1,\dots,n\}$ specifies the mode occupied by the $\alpha^\text{th}$ particle. Since the constituents are identical, the ordering in $\Vec{d}(\Vec{r})$ is irrelevant and here assumed to be given in increasing order of the spatial modes.
    Let us now consider a permutation $\mathcal{P}$ of the $n$ spatial modes which leaves $\psin$ invariant except for a real phase $\varphi$, that is,
    \begin{equation}
    \label{invariance}
        \psin\xrightarrow{\mathcal{P}}e^{i\varphi}\psin.
    \end{equation}
    Notice that $\mathcal{P}$ leaves the internal degree of freedom unaffected. We proceed by computing the eigenvectors of $\mathcal{P}$ and the related eigenvalues $\lambda_1,\lambda_2,\dots,\lambda_n$. Arranging the eigenvectors as columns, we build the matrix $A$ and the unitary evolution matrix $U=A\Sigma$, where $\Sigma$ is an arbitrary diagonal unitary matrix accounting for eventual local phase operations on the output modes. We then focus on the output distribution given by the mode occupation list $\Vec{s}$ and the related mode assignment list $\Vec{d}(\Vec{s})$. Finally, we build the vector $\Vec{\Lambda}(\Vec{s}):=(\lambda_{d_1(\Vec{s})},\dots,\lambda_{d_N(\Vec{s})})$. The suppression law derived in [Phys. Rev. A 97, 062116 (2018)] states that the probability of getting the output distribution $\Vec{s}$ by evolving the input distribution $\Vec{r}$ via $U$ is zero if
    \begin{equation}
    \label{suppression}
        \Pi_{\alpha=1}^{N}\,\Lambda_\alpha(\Vec{s})\neq e^{i\varphi}.
    \end{equation}
    In particular, we notice that Eq.\tilde\eqref{suppression}:
    \begin{enumerate}
        \item
        depends on the input distribution $\Vec{r}$ and the internal input configuration $\Omega_\text{in}=(\ket{I_1},\dots,\ket{I_N})$ characterizing $\psin$ by means of $\Vec{\Lambda}(\Vec{s})$, which is given by the eigenvalues of the permutation $\mathcal{P}$ which satisfies Eq.\tilde\eqref{invariance};

        \item
        depends on the output distribution $\Vec{s}$ via $\Vec{\Lambda}(\Vec{s})$;

        \item
        does not depend on the internal output configuration $\Omega_\text{out}$;

        \item 
        provides a \textit{necessary}, but \textit{not sufficient} condition to obtain the distribution $\Vec{s}$ evolving $\psin$ via $U$, that is, $\Pi_{\alpha=1}^{N}\Lambda_\alpha(\Vec{s})= e^{i\varphi}$.
    \end{enumerate}
    Since we are interested in the suppression law for anti-bunching, we set $\Vec{s}=(\underbrace{1,1,\dots,1}_{N\text{ times}})$ and $\Vec{d}(\Vec{s})=(1,2,\dots,N)$, obtaining $\Vec{\Lambda}(\Vec{s})=(\lambda_1,\lambda_2,\dots,\lambda_N)$ for which Eq.\tilde\eqref{suppression} returns
    \begin{equation}
    \label{generalanti-bunchinglaw}
        \prod_{j=1}^N\lambda_j=e^{i\varphi}.
    \end{equation}

    In the main text, we consider the unitary evolution matrix
    \begin{equation}
        \label{fourier_supp}
            U_N=\frac{1}{\sqrt N}
            \begin{pmatrix}
                1 & 1 & 1 & \dots & 1\\
                1 & \omega & \omega^2 & \dots & \omega^{N-1}\\
                1 & \omega^2 & \omega^4 & \dots & \omega^{2(N-1)}\\
                \vdots & \vdots & \vdots & \vdots & \vdots \\
                1 & \omega^{N-1} & \omega^{2(N-1)} & \dots & \omega^{(N-1)(N-1)}
            \end{pmatrix},
        \end{equation}
        with $\omega=e^{2\pi i/N}$.
        Its columns are the eigenvectors of the cyclic permutation $\cypermut$, whose eigenvalues are
        \begin{equation}
        \label{eigenvalues}
            \lambda_j=\omega^{1-j}=e^{-\frac{2\pi i}{N}(1-j)},
            \quad
            j=1,\dots,N.
        \end{equation}
        The LHS of Eq.\tilde\eqref{generalanti-bunchinglaw} is equal to $\prod_{j=1}^N\omega^{1-j}=e^{{\frac{2\pi i}{N}}\sum_{k=1}^N (1-k)}=(-1)^{N-1}$, so that a necessary condition for a state $\psin$ to anti-bunch under $U_N$ is given by Eq.\tilde\eqref{symmetrycond_supp}.

    \section{Supplemental Note II}
    Here we provide a proof of the relation
    \begin{equation}
    \label{singletprojdecompo}
        \prod_{j=N}^2 P_{E_{\left((-1)^{j-1}\right)}(\pi_{(1,\ldots,j)})}
        =\frac{1}{N!}\sum_{\pi \in S_N} \text{sgn}(\pi)\,\pi
    \end{equation}
    reported in the main text.
    
    Eq.\tilde\eqref{singletprojdecompo} holds for $N=2$; indeed, from the definition
    \begin{equation}
    \label{enprojector_supp}
        \projen:=\frac{1}{N}\sum_{k=1}^{N} \left[(-1)^{N-1}\,\cypermut\right]^k
    \end{equation}
    it follows that
    \[
        P_{E_{-1}(\pi_{(1,2)})}=\frac{1}{2}\left(\id-\pi_{(1,2)}\right)=\frac{1}{2}\sum_{\pi \in S_2} \text{sgn}(\pi)\,\pi.
    \]

    Let us now assume that  Eq.\tilde\eqref{singletprojdecompo} also holds for all $n<N$.
   We have
    \begin{widetext}
        \[
            \begin{aligned}
                \prod_{j=N}^2& P_{E_{((-1)^{j-1})}(\pi(1,\ldots,j))}
                =\projen\left[\prod_{j={N-1}}^{2} P_{E_{\left((-1)^{j-1}\right)}(\pi_{(1,\ldots,j)})}\right]\\
                &=\frac{1}{N}\sum_{k=1}^{N} \Big[(-1)^{N-1}\,\cypermut\Big]^k\,\left[\frac{1}{(N-1)!}\sum_{\pi\in S_{N-1}}\text{sgn}(\pi)\pi\right]\\
                &=
                \frac{1}{N!}\sum_{k=1}^{N} \Big[\underbrace{(-1)^{N-1}}_{\sign{\cypermut}}\,\cypermut\Big]^k\,\left[\sum_{\pi\in S_{N-1}} \text{sgn}(\pi)\pi\right]
                =\sum_{k=1}^{N}\,\sum_{\pi\in S_{N-1}}\,
                \sign{\cypermut^k \pi}\,\cypermut^k\,\pi\\
                &=\frac{1}{N!}\sum_{\pi \in S_N} \text{sgn}(\pi)\,\pi.
            \end{aligned}
        \]
    \end{widetext}
    The last step follows from the fact that both the symmetric group $S_{N-1}$ and the cyclic group $\langle \cypermut\rangle$ are subgroups of $S_N$ and
    \[
        \begin{aligned}
            |S_{N-1} \langle \cypermut\rangle| &= \frac{ |S_{N-1}||\langle \cypermut\rangle|}{|S_{N-1}\cap\langle \cypermut\rangle|}\\
            &=(N-1)! N = |S_N|,
        \end{aligned}
    \]
    where $S_{N-1}\cap\langle \cypermut\rangle=\{1_N\}$ is the identity permutation of $N$ elements.

    \section{Supplemental note III}
    Here we demonstrate that any state $\ket{\Psi}$ from the totally antisymmetric subspace given by the projector $P_{A_N^d}=\frac{1}{N!}\sum_{\pi \in S_N} \text{sgn}(\pi)\,\pi$ with $d\geq N$ is invariant (up to a global phase) under the action of an arbitrary unitary followed by a coincidence measurement on the output modes, that is,
    \begin{equation}
    \label{singlet_invariance}
        C_N U \ket{\Psi} = e^{i \phi} \ket{\Psi}.
    \end{equation}

    A generic state $\ket{\Psi}$ in the totally antisymmetric subspace given by $P_{A_N^d}$ can be written as
    \begin{equation}
    \label{general_antisym_state}
        \ket{\Psi}=
        \sum_{\mathcal{S}\in \mathcal{P}_N([d])}
        c_{\mathcal{S}}
        \ket{A_N^\mathcal{S}},
    \end{equation}
    where $\mathcal{P}_N([d])$ is the family of sets of cardinality $N$ over the set $[d]=\{1,\dots,d\}$, $\ket{A_N^\mathcal{S}}$ denotes an $N$-partite $N$-level singlet state corresponding to the choice $\mathcal{S}\in \mathcal{P}_N([d])$ of $N$ out of $d$ levels, and $c_{\mathcal{S}}$ are coefficients such that $\sum_{\mathcal{S}\in\mathcal{P}_N([d])}|c_{\mathcal{S}}|^2=1$.
    Each generalized singlet state $\ket{A_N^S}$ associated to the choice $\mathcal{S}=\{s_1,s_2,\dots,s_N\}$ can be written in terms of the determinant of a matrix of creation operators as
    \begin{equation}
    \label{singlet_as_determinant}
        \ket{A_N^\mathcal{S}}
        =\frac{1}{\sqrt{N!}}\det{\mathsf{A}_\mathcal{S}}\ket{0},
    \end{equation}
    where
    \begin{equation}
        \mathsf{A}_\mathcal{S}=\begin{pmatrix}
            a^\dagger_{1,s_1} & a^\dagger_{1,s_2} & \ldots & a^\dagger_{1,s_N}\\ 
            a^\dagger_{2,s_1} & a^\dagger_{2,s_2} & \ldots & a^\dagger_{2,s_N}\\
            \vdots&\vdots &\vdots &\vdots\\
            a^\dagger_{N,s_1} & a^\dagger_{N,s_2} &\ldots  & a^\dagger_{N,s_N}
        \end{pmatrix}.
    \end{equation}
    A generic unitary operator $U$ acting on the $N$ spatial modes transforms the creation operators into $u^\dagger_{k,m}=\sum_{\ell=1}^N (U)_{k,\ell}\,a^\dagger_{\ell,m}$, so that
    \begin{equation}
        U|A_N^\mathcal{S}\rangle = \frac{1}{\sqrt{N!}}\det (U \mathsf{A}_\mathcal{S})\, |0\rangle.
    \end{equation}
    Since $\det (U \mathsf{A}_\mathcal{S})=\det U \det \mathsf{A}_\mathcal{S} = e^{i\theta} \det \mathsf{A}_\mathcal{S}$ for some real $\theta$, we get that $U|A_N^\mathcal{S}\rangle=e^{i\theta}|A_N^\mathcal{S}\rangle$. Therefore, from Eq.\tilde\eqref{general_antisym_state} it follows
    \begin{equation}
        \begin{aligned}
            U\ket{\Psi}
            &=\sum_{\mathcal{S}\in \mathcal{P}_N([d])}
            c_{\mathcal{S}}\,
            U\ket{A_N^\mathcal{S}}=e^{i \phi}\ket{\Psi}.
        \end{aligned}
    \end{equation}
    This means that $\ket{\Psi}$ is, up to a global phase, invariant under any unitary operator acting on the $N$ spatial modes. Eq.\tilde\eqref{singlet_invariance} follows from the fact that each $|A_N^\mathcal{S}\rangle$ is a state of $N$ particles occupying $N$ distinct spatial modes, so that it is left invariant by a QND coincidence measurement on the output modes: $C_N |A_N^\mathcal{S}\rangle=|A_N^\mathcal{S}\rangle$.

    Notice that the invariance of the totally antisymmetric state under arbitrary unitaries ensures that our scheme can use any unitary that leads to the same suppression laws as $U_N$. These include the unitaries $U_N'$ resulting from the application of local phases to the input/output modes of the Fourier multiport, $U'=D\,U_N\,D'$, where $D,\,D'$ are diagonal unitary matrices.
    
\end{document}